\documentstyle[12pt]{article}

\def\nnn{\hfill \nonumber \\}
\def\Xb{{\bar X}}

\def\Ab{\bar A}
\def\Bb{\bar B}
\def\Cb{\bar C}

\def\Mb{\bar M}
\def\Ub{{\bar U}}

\def\Ahb{\overline{\widehat A}}
\def\Bhb{\overline {\widehat B}}
\def\Chb{\overline {\widehat C}}

\def\Lhb{\overline {\widehat L}}

\def\Abh{\widehat {\overline A}}
\def\Bbh{\widehat {\overline B}}
\def\Cbh{\widehat {\overline C}}

\def\Mbh{\widehat {\overline M}}
\def\zb{\bar z}

\def\fb{\bar f}

\def\1b{\bar 1}
\def\nb{\bar n}

\def\Cb{\bar C}

\def\phib{\bar \phi}

\def\Au{\underline A}
\def\Bu{\underline B}
\def\Au{\underline A}

\def\0b{\bar 0}
\def\lambdab{\bar \lambda}

\def\Mb{\bar M}

\def\lambdab{\bar \lambda}

\def\beq{\begin{equation}}
\def\eeq{\end{equation}}
\def\beqa{\begin{eqnarray}}
\def\eeqa{\end{eqnarray}}

\begin{document}
\begin{titlepage}
\nopagebreak
\begin{flushright}

LPTENS--93/15 \\
 UT-639
\\
hep-th/9304132
 \\
                April   1993
\end{flushright}

\vglue 1  true cm
\begin{center}
{\large\bf
LIGHT-CONE PARAMETRIZATIONS \\
  FOR K\"AHLER MANIFOLDS}

\vglue 1  true cm
{\bf Jean-Loup~GERVAIS}\\
{\footnotesize Laboratoire de Physique Th\'eorique de
l'\'Ecole Normale Sup\'erieure\footnote{Unit\'e Propre du
Centre National de la Recherche Scientifique,
associ\'ee \`a l'\'Ecole Normale Sup\'erieure et \`a l'Universit\'e
de Paris-Sud.},\\
24 rue Lhomond, 75231 Paris CEDEX 05, ~France.
\\
and
} \\
{\bf Yutaka~MATSUO} \\
{\footnotesize Department of Physics, University of Tokyo \\
Hongo 7-3-1, Bunkyo-ku, Tokyo, Japan.}
\end{center}

\vfill
\begin{abstract}
\baselineskip .4 true cm
\noindent
{\footnotesize It is shown that,  for any K\"ahler manifold, there exist
parametrizations such that the metric takes  a  block-form
identical to  the
light-cone metric
 introduced by Polyakov for two-dimensional gravity.
Besides its possible relevence for various aspects of K\"ahlerian
geometry, this fact allows us  to change gauge in  W gravities,
and explicitly go  from the   conformal (Toda) gauge   to
the light-cone  gauge using the W-geometry we proposed earlier
(this will be   discussed  in detail in a forthcoming article).
}

\end{abstract}
\vfill
\end{titlepage}

It is hardly necessary to stress  the importance of K\"ahler
manifolds. They arise in various important problems of theoretical
physics and mathematics. For instance, K\"ahler
geometries play
an important role in supergravity theories and string
compactification\cite{GSW}\cite{BW}.
For us, the motivation to study K\"ahler manifolds  is our
recent works\cite{GM1}\cite{GM2},
 where they were connected with $W$-geometries. In particular, it was
shown in ref.\cite{GM2} that the  classical solutions of
 $A_n$-$W$ gravity  in the conformal gauge,
are    in a  one-to-one correspondence with holomorphic surfaces
in the complex projective space $CP^n$,  which is the standard
non-trivial example
of K\"ahler manifolds.  On the other hand, $W$ gravities have also
been studied\cite{M1}\cite{LC}\cite{SS}
in a light-cone approach\footnote{it is
also called chiral gauge, but for us this terminology would lead to
confusion, so that we do not use it.}  similar to the
one originally introduced  for two-dimensional gravity\cite{P1}.
So far these two gauges had not been connected. In our study of
$W$-geometries, we came across a general result about
K\"ahler manifolds  which is the point of this letter: starting
from any set of coordinates   of the usual type,  there exist
reparametrizations such that the metric takes  a block-form  identical
to the light-cone metric of two-dimensional gravity introduced
in ref.\cite{P1}.
This results allows us to connect light-cone and
conformal descriptions of W gravities, as we will show in
details in a forthcoming paper. It is, however, of a more general
interest, and we present it separately in this letter.

We shall deal with an arbitrary
K\"ahler manifold ${\cal M}$ of real dimension $2n$.
  By definition
there exists  a special  class of coordinates
 $X^A$, $X^{\Ab}$,
$1\leq A,\, \Ab \leq  n$, such that the only components of the
metric are $G_{A \, \Bb}=G_{\Bb\, A}$, and
\beq
 G_{A \, \Bb}=\partial_A \partial_{\Bb} K,
\label{1}
\eeq
 where
$K$ is the K\"ahler potential. In general we denote the
differential operators $\partial/\partial X^A$, and
 $\partial/\partial X^{\Bb}$ by $\partial_A$, and  $\partial_{\Bb}$.
These coordinates   will be called conformal
 since they appear in the geometrical description
of $W$  gravity in the conformal gauge.
They will be collectively denoted as $X^{\Au}$, with
$1\leq \Au \leq 2n$.
Our basic  point is the definition of another set of prefered
 coordinates denoted $U^{\Au}$
such that the new metric tensor denoted $H_{\Au \, \Bu}$
takes a form which is standard in the light-cone approach
to $W$-gravity. The $U^{\Au}$ will be called light-cone
coordinates.
 In the same way as the
conformal coordinates, they are
split into two sets denoted $U^A$, and $U^{\Ab}$,
respectively. The change of coordinates is of the form
\beq
U^{\Ab}=
U^{\Ab}(X^1,\, \cdots,\, X^n; X^{\1b} \, \cdots,\,  X^{\nb} ), \quad
 U^{A}=X^A.
\label{2}
\eeq
The functions  $U^{\Ab}(X; \Xb)$ will be determined next, in
such a way that
 the light-cone metric takes the form
$$
H_{\Ab \, \Bb}=0,  \quad H_{A \, B} =2 h_{A \, B}
$$
\beq
H_{A \, \Bb}= H_{\Bb \, A} =
\delta_{A, \, \Bb},\hfill
\label{3}
\eeq
where $h_{A \, B}$ will be related to the K\"ahler potential.
Before going on, let us remark that the determinant of $H_{\Au \Bu} $ is
equal to $-1$, and that
its   inverse $H^{\Au \, \Bu}$ is given by
$$
H^{A \, B}=0, \quad H^{\Ab \, \Bb}=
-2 h_{A \, B}
$$
\beq
H^{A \, \Bb}= H^{\Bb \, A} =
\delta_{A, \, \Bb}.\hfill
\label{4}
\eeq
On the contrary, no general form may be given for the determinant of $G$
or its inverse. This is a very nice point of the
light-cone parametrization\footnote{inverting the metric tensor
is often a pain in the neck.}.  Going back to our main line, one finds,
by standard computations,  that  conditions Eqs.\ref{3}
are fulfilled if one has
\beq
G_{A \, \Bb}=H_{A \, \Cb} \partial_{\Bb}U^{\Cb}
\label{5}
\eeq
\beq
2h_{A \, B}=-H_{B \, \Cb} \partial_A U^{\Cb}
-H_{A \, \Cb} \partial_B U^{\Cb}
\label{6}
\eeq
These equations
are  easily solved  by using the K\"ahler potential,
obtaining
\beq
U^{\Ab}=H^{\Ab \, C} \partial_C K,
\label{7}
\eeq
\beq
h_{A \, B}=-\partial_A \partial_B K.
\label{8}
\eeq
Thus we reach the important conclusion that, for any
K\"ahlerian manifold there is a choice of coordinates
such that the metric tensor takes the block-form
\beq
H=\left ( \begin{array}{cc}
2h&1\\
1&0
\end{array} \right )
\label{9}
\eeq
which is the same as the one  introduced by Polyakov
to describe  2D gravity in the light-cone gauge.

In the same way as for two-dimensional gravity in the light-cone gauge,
 it is convenient to develop  the tensor calculus
appropriate to this light-cone parametrization.   For this,
it is useful to introduce once for all a special
notation. In many cases, we have to contract indices,
not with the full metric tensor $H$, but only with its
off-diagonal part $H_{A \, \Bb}=\delta_{A \, \Bb}$ or
with $H^{A \, \Bb}=\delta_{A \, \Bb}$. In practice this
leads to equate  numerical values  of  bar and unbar
indices leading to confusions between bar and unbar
components. We shall use the following convention.
Consider a covariant vector $V^{\Au}$. The index is
raised and lowered using the usual convention:
$V_{\Bu}=H_{\Bu \, \Au} V^{\Au}$, and so on. However,
it is convenient to define
\beqa
V^{\Ahb }&\equiv& H_{A \, \Bb } V^{ \Bb}, \quad
V^{\Abh }\equiv H_{\Ab \, B} V^{B}; \nnn
V_{\Ahb }&\equiv& H^{A \, \Bb} V_{ \Bb}, \quad
V_{\Abh }\equiv H^{\Ab \, B} V_{B}.
\label{10}
\eeqa
The basic idea of this notation is that this contraction
sets  numerical values of indices equal but does not change
their tensorial character. Thus, for instance,  the left
equation of the first line above contains  the
upper bar-component of $V$ with the numerical value of
the index set equal to $A$. As an example of this rule, the
correspondence between covariant and contravariant vectors
explicitly reads
$$
V_A=V^{\Ahb}+2h_{AB} V^B,\quad
V_{\Ab}=V^{\Abh}
$$
\beq
V^{\Ab}=V_{\Abh}-2h_{\Abh  \Bbh} V_{\Bb}, \quad
V^{A}=V_{\Ahb}.
\label{11}
\eeq
Another example is that Eq.\ref{7} becomes
\beq
U^{\Ahb}=\partial_A K
\label{12}
\eeq
 Concerning the Christoffel symbols denoted
$\Gamma$, the usual text-book calculation
gives in  the present case
\beqa
\Gamma^C_{AB}&=&-D_{\Chb} h_{AB} \hfill \nonumber \\
\Gamma^{\Cb}_{AB}
&=& D_{A}h_{\Cbh B}+D_B h_{A \Cbh}
-D_{\Cbh} h_{AB} +2h_{\Cbh \Mbh} D_{\Mb}h^{AB} \nnn
\Gamma_{A \, \Bb}^C&=& D_{\Bb} h_{A \Cbh} \nnn
\Gamma_{A \Bb}^C&=&\Gamma_{\Ab \Bb}^C
=\Gamma_{\Ab \Bb}^{\Cb}=0.
\label{13}
\eeqa
The partial derivatives with respect to $U^A$  and $U^{\Ab}$
are denoted by $D_A$ and $D_{\Ab}$.

Our next topic is to look for coordinate reparametrizations
that leave the light-cone form Eq.\ref{3} invariant. Under the
infinitesimal change of coordinate $U^{\Au}={\widetilde U}^{\Au}
+\epsilon^{\Au}$, the variation of the metric is
\[
\delta H_{\Au \, \Bu} =
{\cal D}_{\Au} \epsilon _{\Bu} +
{\cal D}_{\Bu} \epsilon_{\Au}
\]
where ${\cal D}$ denotes covariant derivatives with respect to
the Christoffel symbols Eqs.\ref{13}. Imposing first that
$H_{A \, \Bb}$ and $H_{\Ab \, \Bb}$   remain unchanged  gives
\beq
\delta H_{A \, \Bb}=0=D_A \epsilon_{\Bb}
+D_{\Bb} \epsilon_A -2 (D_{\Bb} h_{A \, \Cbh})\>  \epsilon_{\Cb}
\label{14}
\eeq
\beq
\delta H_{\Ab \, \Bb}=0
=D_{\Ab} \epsilon_{\Bb}+D_{\Bb} \epsilon_{\Ab}
\label{15}
\eeq
Assuming that these conditions hold, we consider the variation
of $H_{AB}$, that is
\[
\delta H_{AB}=
D_A\epsilon_B+D_B\epsilon_A-2\Gamma_{AB}^C\epsilon_C
-2 \Gamma_{AB}^{\Cb}\epsilon_{\Cb}
\]
Next we show that, in close analogy  with  2D gravity, this may
be rewritten  solely as a function of
$\epsilon^{\Chb}+h_{CL}\epsilon^L$.
This is achieved  as follows: using Eq.\ref{11} we write
\[
\delta H_{AB}=
\left [ D_A \delta_{C B} +D_B \delta_{C A} -2\Gamma_{AB}^C
\right ] (\epsilon^{\Chb}+2 h_{CL} \epsilon^L)
 -2 \Gamma_{AB}^{\Cb}\epsilon_{\Cb}.
\]
In the first term, we change the coefficient of $h_{CL}$
from $2$ to $1$,
and introduce $\tilde \delta H_{AB} $ such that
\[
\delta H_{AB}
=\left [ D_A \delta_{C B} +D_B \delta_{C A} -2\Gamma_{AB}^C
\right ] (\epsilon^{\Chb}+ h_{CL} \epsilon^L) + \tilde \delta H_{AB}.
\]
\[
\tilde \delta H_{AB}=D_A( h_{BL}\epsilon^L)
+D_B (h_{AL} \epsilon^L)  -2\Gamma_{AB}^C h_{CL} \epsilon^L
-2 \Gamma_{AB}^{\Cb}\epsilon_{\Cb}
\]
Next, we make use of Eqs.\ref{14}, and \ref{15}, and  derive
\[
(h_{BC} D_A+h_{AC} D_B) \epsilon^C=
\]
\[
-h_{BC}D_{\Chb} \left [ \epsilon ^{\Ahb}+h_{A L} \epsilon^L\right ]
-h_{AC}D_{\Chb} \left [ \epsilon ^{\Bhb}+h_{B L} \epsilon^L\right ]
\]
\[
+h_{BC}  D_{\Chb}(h_{AL}) \epsilon^L
+h_{AC}  D_{\Chb}(h_{BL})  \epsilon^L.
\]
\[
\tilde \delta H_{AB}=
-\left [ h_{CB} \delta_{AL} D_{\Chb} +h_{CA} \delta_{BL} D_{\Chb}
\right ](\epsilon^{\Chb}+ h_{CL} \epsilon^L)  + V_C \> \epsilon^C
\]
\[
V_C= h_{BL}D_{\Lhb} h_{AC} + h_{AL}D_{\Lhb} h_{BC}
+D_A h_{BC}+D_B h_{AC} -2 \Gamma^L_{AB}h_{LC} -2 \Gamma^{\Chb}_{AB}.
\]
Finally, one verifies that $V_C$ is actually equal to zero. This is a
consequence of the K\"ahler condition, which allows us to transform
the expression of $\Gamma^{\Chb}_{AB}$. Indeed, using the fact that
$\partial_A=D_A-h_{AL}D_{\Lhb}$, one sees that the expression
for $\Gamma^{\Chb}_{AB}$ given in Eq.\ref{13} is equivalent to
\beq
\Gamma^{\Chb}_{AB}=
{1\over 2} \left ( D_{A} h_{BC} +D_B h_{AC} +h_{AL} D_{\Lhb} h_{BC}
+h_{BL} D_{\Lhb} h_{AC} \right ) +h_{CL} D_{\Lhb} h_{AB}.
\label{16}
\eeq
Substituting this last expression into the above formula for $V_C$ one
finds that there is a complete cancellation between the last term and
the others. Collecting  the remaining pieces, one arrives at the
 formula
\[
\delta H_{AB}=
\]
\beq
\left [ D_A \delta_{C B} +D_B \delta_{C A}
-(h_{B\Mbh} \delta_{AC}+h_{A\Mbh} \delta_{BC})D_{\Mb}
+2D_{\Chb} h_{AB}
\right ] (\epsilon^{\Chb}+ h_{CL} \epsilon^L).
\label{17}
\eeq
As already announced, it only involves the quantities
\beq
v_C\equiv \epsilon^{\Chb}+h_{CL}\epsilon^L=
\epsilon_C-h_{CL}\epsilon^L.
\label{18}
\eeq
It may be written compactly as
\beq
\delta H_{AB}= (\nabla_A v)_B+(\nabla_B v)_A
\label{19}
\eeq
\beq
(\nabla_A v)_B\equiv (D_A-h_{A\Mbh} D_{\Mb}) v_B +
(D_{\Chb} h_{AB})\>  v_C.
\label{20}
\eeq
This generalizes a basic formula of 2D gravity in the
light-cone gauge which reads, with  standard  notations,
$\delta h=(\partial_+-h\partial_-+\partial_+h)(\epsilon^++h\epsilon^-)$.

The usual K\"ahler formulation is obviously covariant under
holomorphic change of coordinates $X^A\to X^A( \widetilde X^1, \cdots,
\widetilde X^n)$,
$X^{\Ab}\to X^{\Ab}( \widetilde X^{\1b}, \cdots,
\widetilde X^{\nb})$. Then $K\to \widetilde K$
such that $K(X,\Xb)= {\widetilde K} (\widetilde X, {\widetilde {\Xb}})$.
Using the change to
light-cone coordinates just displayed, this gives
examples of transformations that leave
the light-cone form invariant. Consider
the infinitesimal holomorphic transformation
\beq
X^A={\widetilde X}^A-\eta^A(X),\quad X^{\Ab}=
{\widetilde X}^{\Ab}-\eta^{\Ab}(\Xb).
\label{21}
\eeq
It immediately follows from Eq.\ref{12} that
\beq
\delta U^{\Ahb}=-(\partial_A \eta^C)\partial_C K
=-U^{\Chb}D_A\eta^C,
\label{22}
\eeq
and Eq.\ref{2} directly shows that
\beq
\delta U^{A}=\eta^A.
\label{23}
\eeq
Thus the antiholomorphic part $\eta^{\Ab}(\Xb)$ does not act, and
the variation of the $\Ub$ coordinates is linear in $\Ub$.
One sees that,  when one goes to the light-cone coordinates, the
covariance under anti-holomorphic transformations is lost. On the
other hand, other transformations appear.
First, the light-cone formulation is not
invariant under the change of the K\"ahler potential
\beq
\delta K= \phi(X)+\phib(\Xb)
\label{24}
\eeq
that leave the original metric Eq.\ref{1} invariant.
A simple calculation shows that
\beq
\delta U^{\Ahb}=-D_{A} \phi,\quad \delta U^{A}=0
\label{25}
\eeq
This,  give another  example of
transformations of the type Eq.\ref{20}.
Second, we did not write down the most general solution of
Eqs.\ref{5}, and \ref{6}. It is given by
\beq
U^{\Ahb}=\partial_A K+\Omega_{A}(X),\quad
h_{A \, B}=-\partial_A \partial_B K-{1\over 2} (\partial_A \Omega_{B}(X)+
\partial_B \Omega_{A}(X)).
\label{12'}
\eeq
$\Omega_{A}$ are arbitrary
functions of $X^1,\, \cdots , \, X^n$, or,
equivalently, of $U^1,\, \cdots , \, U^n$. Changing $\Omega$
gives another  set of transformations that  leave the physics
invariant, and give examples of Eq.\ref{19}, and \ref{20}.

The change of coordinates just described came out in our study of
W gravity as follows. As already recalled, we showed
in ref.\cite{GM2},  that the
$A_n$-W--geometry corresponds  to the embedding of holomorphic
two-dimensional surfaces in $CP^n$. These ($W$) surfaces are
specified by
embedding equations  of the from $X^A=f^A(z)$,
$\Xb^{\Ab}=\fb^{\Ab}(\zb)$,
where $z$, and $\zb$ are the two surface-parameters. The fact that
they are functions of a single  variable is equivalent to
the Toda field-equations, so that this describes W gravity in
the conformal gauge.
These functions have a natural extension to $CP^n$ using the
higher variables $z^{(k)}$, $\zb^{(k)}$
 of the Toda hierarchy of integrable flows, and this
provides a local parametrization of $CP^n$. The original variables
$z$ and $\zb$ are identified with $z^{(1)}$, $\zb^{(1)}$, respectively.
 For  the embedding
 functions the extension is such that they become functions of
half  of the variables noted
$f^A([z])=f^A(z^{(0)}, \cdots z^{(n)})$, and
$\fb^{\Ab}([\zb])=\fb^{\Ab}( \zb^{(0)}, \cdots \zb^{(n)})$ such that
\beq
{\partial f^A([z]) \over \partial z^{(k)}}  =
{\partial^k f^A([z]) \over (\partial z)^k}, \quad  \quad
{\partial \fb^{\Ab}([\zb]) \over \partial \zb^{(k)}}  =
{\partial^k \fb^{\Ab}([\zb]) \over (\partial \zb)^k}
\label{26}
\eeq
 One  main virtue of the coordinates
$z^{(k)}$, $\zb^{(k)}$ is that,  due to the last equations,
 higher derivaties in $z$ and $\zb$ are changed
to first-order ones, and this is how our geometrical
scheme gets rid of the  troublesome higher derivatives of the
usual approachs. So far this is only for the conformal gauge.
Our  new result  is that,
{\bf performing the change of coordinates Eqs.\ref{2},
\ref{7}, \ref{8} in the target space $CP^n$, allows us  to go from
W-gravity in the conformal gauge to $W$-gravity
in the light-cone gauge}.
{}From this viewpoint, the transformations Eqs.\ref{19} -- \ref{21}
just display the local gauge group of $W$ gravity in the
light-cone gauge. They  only involve  first order derivatives in
the target space. Higher derivatives appear when the higher coordinates
are eliminated by means of Eq.\ref{26}. This
will be spelled out later on in full details.  Right now we
discuss another general aspect,   inspired by the problem
of W gravity, which is  the existence of a Lax pair in the conformal
gauge, with a vector-potential  related with $h_{AB}$. We denote by
small gamma's  the Christoffel symbols in the conformal parametrization.
As is well known, the only non-zero components are $\gamma_{AB}^C$, and
$\gamma_{\Ab \Bb}^{\Cb}$, so that we have
\[
\partial_M G_{\Ab B}= G_{\Ab C} \gamma_{MB}^C, \qquad
\partial_{\Mb}  G_{\Ab B}=  \gamma_{\Mb \Ab}^{\Cb} G_{\Cb B}.
\]
This may be easily rewritten in a Lax-pair from
\beq
\partial_M G_{\Ab C}= {\cal A}_{(M) \Ab}^{\Bb}G_{\Bb C},
\qquad \partial_{\Mb} G_{\Ab C}=  {\cal A}_{(\Mb) \Ab}^{\Bb}G_{\Bb C},
\label{27}
\eeq
where
\beq
{\cal A}_{(M) \Ab}^{\Bb}=G_{\Ab C} \gamma_{M D}^{C} G^{D \Bb}, \qquad
{\cal A}_{(\Mb) \Ab}^{\Bb}= \gamma_{\Mb \Ab}^{\Bb}.
\label{28}
\eeq
Moreover, it follows from the K\"ahler condition Eq.\ref{1},
that
\beq
{\cal A}_{(M) \Ab}^{\Bb}=-(\partial_{\Ab}  H_{MD})  G^{D \Bb},
\label{29}
\eeq
and this establishes the connection between the present vector
potential and the light-cone metric tensor Eq.\ref{3}.
It is straightforward to verify that the connection $\cal A$ is indeed
flat, so that the last equation does define a Lax pair.
What is the meaning for W gravity ? Consider the components
${\cal A}_{(1)}$, and ${\cal A}_{(\bar 1)}$. We showed in ref.\cite{GM2}
that when one uses the $z^{(k)}$ , and $\zb^{(k)}$, as
homogeneous coordinates for $CP^n$,
 the Christoffel symbols become very simple.  The  connection
${\cal A}_{(\bar 1)}$ is given by
\beq
{\cal A}_{(\bar 1)}= I+\lambdab
\label{30}
\eeq
\beq
I =\left (\begin{array}{ccccc}
0 & 1 & 0 &\cdots & 0 \\
0 & 0 & 1 &\cdots & 0 \\
\vdots & \,&\ddots & \ddots  & \vdots \\
0 & \cdots  & \, &0 & 1 \\
0 & \cdots  & \, &0 & 0\end{array}
\right ), \quad
\lambdab=\left (\begin{array}{ccccc}
0 & 0 & 0 &\cdots & 0 \\
0 & 0 & 0 &\cdots & 0\\
\vdots & \, & \ddots &\,  & \vdots \\
0 & \, & \cdots &0  & 0\\
\lambdab_0 & \lambdab_1  & \cdots &\lambdab_{n-1} &  \lambdab_n
 \end{array}
\right ),
\label{31}
\end{equation}
where the $\lambdab_i$ are related with the $W$ charges. So far we
were considering points in the target space $CP^n$. According to
our previous work\cite{GM2}, the usual two dimensional dynamics
of W gravity is recovered if one  returns
to the $W$ surface by letting $z^{(k)}=0$, and $\zb^{(k)}=0$,
for $k\not =1$, and $z^{(1)}=z$, and $\zb^{(1)}=\zb$. Then the form of
${\cal A}_{(\bar 1)}$ is precisely the one that comes out in the
Drinfeld-Sokolov equation\cite{DS}, in the Hamiltonian
reduction\cite{BO}, and in the generalized Beltrami-differential
approach\cite{BFK}. In particular, this last reference displays   a
connection between the Lax pair just written, and W gravity
in the light-cone gauge.
What we just described  gives the geometrical
origin of this connection, which basically follows from change of
coordinates in the target space $CP^n$. In refs.\cite{SS},
and \cite{BFK}, it is shown that the anomaly equations  of light-cone
W gravity precisely comes  from the zero-curvature conditions
associated
with a Lax pair of the type we just wrote. In our approach they
thus follow  from the K\"ahler condition together with the existence of
a parametrization where the Christoffel symbols take the form
Eq.\ref{31}. There is a subtle difference between the two, however,
since starting from Toda field equations we can  only get
conformally invariant results, contrary to the work of ref.\cite{BFK}.
This point will be discussed in detail later on.

The condition on
the Christoffel symbols  is strongly reminiscent of the one that specifies
a particular Toda theory, in the group-algebraic approach\cite{LS1}
-- \cite{LS3}  to
Toda dynamics, and in the conformally reduced WZNW approach\cite{GORS}.
Thus a similar mechanism will probably work for the other
$W$ geometries.

In conclusion, we have displayed a change of coordinates for arbitrary
K\"ahler
geometries that  leads to a metric tensor similar to the one of
the light-cone formulation of W gravities.  These geometries are so
important in various problems, that this result will probably be usefull
beyond the W gravity problems which was the motivation of this work.

\noindent {\bf Acknowledgements}

This work was completed while one
of us (J.-L. G.) was visiting Japan during March 1993. He is
grateful to the Physics Department of Tokyo University, to the
Yukawa Institute of Kyoto, and to the KEK,  for
their warm hospitalities and generous financial supports.
This work is supported in part from
Grant-in-Aid for Scientific Research on Priority Area,
the  Ministry of Education,
Science and Culture, Japan.


\begin{thebibliography}{99}

\bibitem{GSW} M. Green, J. Schwarz, E. Witten: "Superstring theory",
Cambridge Univ. Press (1987)

\bibitem{BW} see e.g. B. de Witt:  ``String and symmetries
91'', {\sl in} Proceeding of the
Stony Brook Meeting, World Scientific ed.; and
refs. therein.

\bibitem{GM1} J.-L. Gervais, Y. Matsuo:
{\sl Phys. Lett.} {\bf 274B} (1992) 309;

\bibitem{GM2} J.-L. Gervais, Y. Matsuo:
Comm. in Math. Phys. {\bf 152} (1993) 317.

\bibitem{M1} Y. Matsuo: Phys. Lett. B {\bf 227}
(1989) 209.

\bibitem{LC} see, for instance,  the related
reviews in  ``String and symmetries
91'', {\sl in} Proceeding of the
Stony Brook Meeting, World Scientific ed.

\bibitem{SS} G. Sotkov, M. Stashnikov: Nucl. Phys. B
{\bf 356} (1991) 439; G. Sotkov, M. Stashnikov,
M. Zhu: Nucl. Phys. B
{\bf 356} (1991) 245.


\bibitem{P1} A. Polyakov: Mod. Phys. Lett. A {\bf 2, 11}
(1987) 893.

\bibitem{DS}
V.G. Drinfeld and V.V. Sokolov: {\sl Journ. Sov. Math.}
{\bf 30} (1985) 1975.

\bibitem{BO} M. Bershadsky, H. Ooguri: Comm. Math. Phys.
{\bf 126} (1989) 49.

\bibitem{BFK}
A. Bilal, V.V. Fock and I.I. Kogan:
{\sl Nucl. Phys.} {\bf B359} (1991) 635.

\bibitem{LS1}
A.~N.~Leznov, M.~V.~Saveliev: Phys. Lett. {\bf B79} (1978) 294;
Lett. Math. Phys. {\bf 3} (1979) 207; Comm. Math.
Phys. {\bf 74}  (1980) 111.
\bibitem{LS2}
A.~N.~Leznov, M.~V.~Saveliev:
Lett. Math. Phys. {\bf 6}  (1982) 505; Comm. Math.
Phys. {\bf 89}  (1983) 59.
\bibitem{LS3}
A.~N.~Leznov, M.~V.~Saveliev:
{\it Group-Theoretical Methods for Integration
of Nonlinear Dynamical Systems.} Progress
in Physics v.~15, Birkhauser-Verlag,
1992.

\bibitem{GORS} see J.-L. Gervais,
L. O'Raifeartaigh, A. Razumov, M. Saveliev:
Phys. Lett. {\bf B301} (1993) 401, and references therein.






 \end{thebibliography}
\end{document}